\documentstyle[prd,aps,floats]{revtex}
\begin{document}
\draft

\input epsf \renewcommand{\topfraction}{0.8}
\twocolumn[\hsize\textwidth\columnwidth\hsize\csname
@twocolumnfalse\endcsname

\newcommand{\be}{\begin{equation}}
\newcommand{\ee}{\end{equation}}
\newcommand{\bea}{\begin{eqnarray}}
\newcommand{\eea}{\end{eqnarray}}
\newcommand{\kma}{\; ,}
\newcommand{\pkt}{\; .}
\newcommand{\calO}{{\cal O}}

%\preprint{DO-TH 01/07, LA-01-2141, SUSX-TH/01-019}
\title{CMB constraints on non-thermal leptogenesis}
\author{Anupam Mazumdar}
\address{CHEP, McGill University, Montr\'eal, QC, H3A 2T8, Canada}
\maketitle
\begin{abstract}
Leptogenesis is at the heart of particle cosmology which requires
physics beyond the Standard Model. There are two possibilities of
realizing leptogenesis; thermal and non-thermal. Both are viable given
the scale of inflation and the constraint on the reheat temperature.
However non-thermal leptogenesis can leave its imprint upon cosmic
micro wave background radiation. In this paper we will discuss
cosmological constraints on non-thermal leptogenesis scenarios within
supersymmetry.
\end{abstract}

\vskip2pc]
%%%%%%%%%%%%%%%%%%%%%%%%%%%%%%%%%%%%%%%%%%%%%%%%%%%%%%%%%%%%%%%%%%%%%%%%%%

\section{Introduction}

Baryogenesis and neutrino oscillations are the two fronts which
naturally evoke physics beyond the electroweak Standard Model
(SM). Both Big bang nucleosynthesis and current WMAP data suggest
that the baryon asymmetry is of the order of one part in
$10^{10}$~\cite{Olive}, while the solar neutrino experiments suggest
$\Delta m^2_{solar}\sim 7\times 10^{-5}~{\rm eV}^2$ with large mixing
angle $\tan^2\theta_{solar}\sim 0.5$~\cite{Fukudas}, and the atmospheric
($\nu_{\mu}-\nu_{\tau}$) oscillations with $\Delta m^2_{atm}\sim
2.5\times 10^{-3}~{\rm eV}^2$ and $\sin^2(2\theta)\simeq 1$
\cite{Fukudaa}. Within the SM all three Sakharov's conditions cannot be
realized at the same time: baryon number violation, $C$ and $CP$
violation, and strong out of equilibrium condition~\cite{Sakharov}. 
Especially the last one is difficult to achieve with the Higgs mass 
constraint $\geq 114$~GeV from the LEP experiment~\cite{LEP}. 
There is a small range of parameter space left within minimal 
supersymmetric Standard Model (MSSM) where the electroweak baryogenesis 
can still work with the lightest stop mass lighter
than the top quark mass~\cite{Carena}. There are currently other
popular schemes of baryogenesis such as MSSM flat direction induced
baryogenesis (for a review see~\cite{Enqvistrep}). An interesting
point of Affleck Dine baryogenesis is that it generates
baryon-isocurvature fluctuations~\cite{McDonald}, which can be
constrained from the cosmic microwave background (CMB) data.

On the other hand the observed light neutrino masses can be obtained
naturally if the Majorana nature of light neutrinos is confirmed 
along with a see saw scale~\cite{Seesaw}. An advantage of this 
is leptogenesis via $L$ or $B-L$ violation and its subsequent 
conversion to baryon asymmetry through active SM sphalerons: 
$10^{12}~{\rm GeV}\geq T\geq 100$~GeV~\cite{Fukugitta}.

It is almost impossible to test leptogenesis in a model independent
way, because of the uncertainties in the scale of leptogenesis and
the appearance of $CP$ phase participating in leptogenesis. Especially 
this phase need not be the same as that of the low energy $CP$ phase 
in the left handed neutrino sector. In a $3\times 3$ scheme, where
there are $3$ heavy right handed and $3$ light neutrinos, there are
$18$ real parameters and $3$ $CP$ violating phases. It has been proven
extremely hard to make any prediction~\cite{Joshipura}. Some progress 
were made with $2$ heavy and $3$ light neutrino species where there 
are now $8$ real parameters and $3$ $CP$ violating phases~\cite{Endoh}.

In this paper we will not follow the conventional argument of testing
leptogenesis via measuring $CP$ phases, but we will look forward to
cosmology and particularly the physics of the cosmic microwave background
radiation. In this regard we need two minimal assumptions: embedding
leptogenesis in a supersymmetric set up and assume inflation. Both
supersymmetry and inflation are necessary in their own rights. The
advantage of supersymmetry is that it naturally provides a scalar
component of the right handed neutrino field, sneutrino. On the other
hand inflation creates a condensate for the sneutrino field with a
non-vanishing vacuum expectation value (vev). If the lightest
sneutrino mass is smaller than the Hubble expansion, then
during inflation the quantum perturbations are stretched outside the
horizon. The fermions also fluctuates during inflation, however they
cannot be treated as a condensate due to lack of large occupation
number. For earlier discussions on sneutrino induced leptogenesis,
see~\cite{Murayama1,Berezhiani,Moroi1}. Our discussion differs from
that of Ref.~\cite{Murayama1,Moroi1} in some respects. We will always
assume that the sneutrino condensate is not responsible for inflation,
e.g. the inflaton energy density dominates over the sneutrino energy density,
we will explain why we require so.

%%%%%%%%%%%%%%%%%%%%%%%%%%%%%%%%%%%%%%%%%%%%%%%%%%%%%%%%%%%%%%%

\section{Thermal vs Non-thermal Leptogenesis}

Leptogenesis can be thermal or non-thermal. In a thermal case the 
asymmetry is given by, for a review see~\cite{Buchmuller},
\begin{equation}
\frac{n_{B}}{s}\approx \frac{8}{15}\frac{\epsilon_{1}}{g_{\ast}}\times\kappa\,,
\end{equation}
where $s$ is the entropy, the numerical factor accounts for the lepton
baryon asymmetry in MSSM with two Higgs doublet, $\epsilon_{1}$ is the
$CP$ asymmetry of the lightest right handed neutrino, and
$g_{\ast}\sim {\cal O}(100)$ is the relativistic degrees of
freedom. $\kappa\sim 10^{-1}-10^{-2}$ is a measure of dilution
estimated numerically by solving the Boltzmann equation for $\Delta
L=1,~\Delta L=2$ washout processes~\cite{Bari}. Therefore yielding
$n_{B}/s\sim (10^{-3}-10^{-4})\epsilon_{1}$.

On the other hand in the non-thermal leptogenesis the net asymmetry
usually depends on the temperature of the decaying particles. If the 
lepton asymmetry is created before or during the inflaton decay, such
that $\Gamma_{N}\geq \Gamma_{X}$, then the net baryon asymmetry is 
given by 
\begin{equation}
\frac{n_{B}}{s} \approx \epsilon_{1}\frac{T_{rh}}{m_{\phi}}\,,
\end{equation}
where $T_{rh}$ is the reheat temperature of the Universe and $m_{\phi}$
is the inflaton mass, the ratio of two arises due to entropy 
generation from the inflaton decay. For an example, for 
$T_{rh}\sim 10^{9}$~GeV, in order not to over produce thermal 
\cite{Ellis} and non-thermal gravitinos~\cite{Maroto}, and 
$m_{\phi}\sim 10^{13}$~GeV, in the case of chaotic inflation, then 
the net baryon asymmetry can be given by $n_{B}/s\sim 10^{-4}\epsilon_{1}$. 
Note that comparatively small $\epsilon_{1}\leq 10^{-6}$ is required 
to yield a net baryon asymmetry.

However if the baryon asymmetry is solely created from the decay of the
right handed neutrinos such that $\Gamma_{N}\leq \Gamma_{X}$, then 
the baryon asymmetry is given by
\begin{equation}
\frac{n_{B}}{s} \approx \epsilon_{1}\gamma\frac{T_{N}}{m_{N}}\,,
\end{equation}
where $m_{N}$ is the right handed neutrino mass and $\gamma$ 
accounts for the possible dilution of the asymmetry due to the entropy
generation during reheating.

The advantages of thermal leptogenesis is that it requires minimal
parameters, just $CP$ asymmetry, e.g. $\epsilon_1$, while non-thermal
leptogenesis undergoes the uncertainties of thermalization.
Nevertheless non-thermal leptogenesis is inevitable in a
supersymmetric context, as we argued earlier. The off-shoot of
inflation is the formation of the lightest sneutrino condensate, and
if this condensate survives, e.g. thermal scattering and evaporation,
then the sneutrino induced leptogenesis is a rather natural phenomenon.

The greatest advantage of non-thermal leptogenesis is that it is testable 
from CMB, because the fluctuations in the sneutrino condensate can be 
transferred into the fluctuations in the baryon asymmetry, which gives
rise to the baryon-isocurvature fluctuations. It is easy to see,
where the reheat temperature obtains spatial fluctuations, the baryon 
asymmetry $\eta =(n_{B}/s)$ also obtains the large scale fluctuations,
\begin{equation} 
\frac{\delta\eta}{\eta} \sim {\cal O}(1) \frac{\delta T_{rh}}{T_{rh}}\,. 
\end{equation}
On contrary thermal leptogenesis can never be tested in this way.
Some critics may ponder on the feasibility of testing non-thermal
leptogenesis via isocurvature fluctuations, because there could be 
many sources generating isocurvature fluctuations in the early 
Universe. The most popular paradigm could be the cold dark matter (CDM)
isocurvature fluctuations, nevertheless within SUSY, thermal generation
of CDM is likely to happen. Here we will rather take an opportunistic 
view point with the possibility of constraining right handed 
neutrino mass scale from CMB, which is an interesting topic in its own right.

%%%%%%%%%%%%%%%%%%%%%%%%%%%%%%%%%%%%%%%%%%%%%%%%%%%%%%%%%%%%%%%%%%

\section{Two models for neutrino masses}

The neutrinos obtain masses via Yukawa couplings from the Higgs 
vev. The Higgs field couples to the inflaton via a superpotential 
term, see for example~\cite{Asaka},
\begin{equation}
\label{superpot}
W=\lambda{\bf X}\Phi\bar\Phi+ g\frac{\Phi\bar\Phi}{M_{\ast}}
{\bf N}{\bf N}+ h{\bf N} {\bf H}_u {\bf L} \,,
\end{equation}
where $X$ is a gauge singlet inflaton, $\Phi$ is the Higgs superfield,
$g,h$ are the Yukawas, ${\bf N}$ is the right handed neutrino
superfield, and $M_{\ast}$ is the fundamental cut off of the theory,
which could be either the string or the Planck scale $M_{p}=2.4\times
10^{18}$~GeV. The right handed neutrinos obtain mass from the vev
$v_{\phi}$ after the end of inflation, which is given by (we will call
this kind of models as type $I$ model),
\begin{equation}
\label{type1}
M_{N}\propto g\frac{v_{\phi}^2}{M_{p}}\,,
\end{equation} 
where $g$ is a $3\times 3$ matrix, here we have ignored the
texture of the right handed neutrino mass matrix and we always 
assume a diagonal basis for the right handed neutrino mass matrix.
The light neutrinos obtain masses via see-saw mechanism
$m_{\nu}\sim m_{D}^2/M_{N}$, where $m_{D}$ is the Dirac mass.
For $v_{\phi}\sim 10^{15}$~GeV, the scale for the right handed 
neutrino masses comes out to be around $10^{12}$~GeV. Note that 
with the above superpotential term the right handed neutrino 
masses are identically zero during inflation, because $X$ being 
an inflaton is only rolling down the potential while the Higgs 
is settled in its minimum, e.g. $\Phi,~\bar\Phi =0$. Inflation 
is supported by the Higgs vev, e.g. $V_{inf}\sim \lambda^2 v_{\phi}^4$.

There is also a non-renormalizable superpotential for the right 
handed neutrino field, which is valid below the $U(1)_{B-L}$ breaking 
scale or the SO(10) breaking scale, $v_{\phi}$,
\begin{equation}
\label{contribute1}
W=\lambda_{1}\frac{N^{n}}{nM_{p}^{n-3}}\,.
\end{equation}
If $R$-parity is conserved then the right handed neutrino direction is
lifted by $n=4$ operator, and $|\lambda_{1}|\leq {\cal O}(1)$ is treated 
as a free parameter. Besides this correction there are various soft 
SUSY breaking terms, such as $(m_{3/2}^2+ C_{N}H^2){\widetilde N}^2$, 
where $m_{3/2}$ is the gravitino mass. The $A$-terms, $(a_1m_{3/2}{\widetilde N}h_{u}{\widetilde l} +a_2H{\widetilde N}h_{u}{\widetilde l})$, and
the non-renormalizable potential for the sneutrino,
$(a_3{H}/{M_{p}^{n-3}}{\widetilde N}^{n}+a_4{m_{3/2}}/{M_{p}^{n-3}}{{\widetilde N}^{n}}+{\rm h.c.})$,
where $|a_i|\sim {\cal O}(1)$ are complex numbers, and tilde denotes 
sparticle. We always consider $n=4$ in our example. The Hubble induced 
$A$-terms and the Hubble induced soft SUSY breaking mass terms are possible 
if the inflaton potential, $v_{\phi}$, arises from the $F$-sector. 
However in the $D$-term inflation case the Hubble induced mass and the 
Hubble induced $A$-term correction does not arise.

Note that if we embed the right handed neutrino sector into a 
gauge group, e.g. $SO(10)$, then we would also expect the D-term 
contributions for $\widetilde N$. However if the vev of $\widetilde N$ 
is less than the $SO(10)$ breaking scale then the D-term contribution 
decouples from the rest of the potential~\cite{Murayama1}.

For the sake of illustration let us consider a D-term inflation.
The scalar potential during inflation is then given by  
\begin{eqnarray}
\label{source}
V\approx \lambda^2v_{\phi}^4+\frac{\lambda^4v_{\phi}^4}{16\pi^2}
\ln\left(\frac{X}{M_{p}}\right)+\lambda_{1}^2\frac{\widetilde N^{6}}{M_{p}^2}
\,.
\end{eqnarray}
The second term in the above potential is the one-loop Coleman-Weinberg
contribution due to SUSY breaking. Note that we have neglected
the soft SUSY breaking contributions here. At sufficiently 
large scales it is the non-renormalizable term dominates 
the potential. For simplicity and for the purpose of illustration 
we assume that the lightest right handed electron sneutrino forms 
a condensate.

In this paper we would like to advocate that the inflaton sector always
dominate the sneutrino sector. There have been some suggestions 
regarding sneutrino dominated inflation, but there are problems associated
with that, e.g. setting up a gauged sector for the sneutrino will be
impossible, especially when the sneutrino vev is larger than
$M_{p}$. In this case the inflaton vev, $X$, can be related to the
number of e-foldings of inflation, ${\cal N}_{e}$, e.g.
\begin{equation}
\label{lim}
X \approx \lambda\frac{M_{p}}{2\pi}\sqrt{{\cal N}_{e}}\,.
\end{equation}
The initial vev of $\widetilde N$ has to be smaller than the critical
value, in order not to ruin the inflation, which is given by
\begin{equation}
\label{critical}
{\widetilde N}_{c}=\sqrt{2}\left(\frac{\lambda}{\sqrt{2}\lambda_1}v_{\phi}^2
M_{p}\right)^{1/3}\,.
\end{equation}
We note that in this class of models the slow roll conditions are 
governed by the inflaton, $\epsilon_{X}\ll |\eta_{X}|\approx M_{p}^2{V(X)^{\prime \prime}}/{V_0}\approx {1}/({2{\cal N}_{e}})\ll 1$,
where $V_0=\lambda^2v_{\phi}^4$. The spectral index is given by
\begin{eqnarray}
n-1&\approx &-3M_{p}^2\frac{V^{\prime}(X)^{2}}{V_{0}^2}+
2M_{p}^2\frac{V^{\prime\prime}(X)}{V_0}-M_{p}^2\frac{V^{\prime}
({\widetilde N})^{2}}{V_{0}^2}\,2\eta_{X}\nonumber\\
&\sim & {1}/{{\cal N}_{e}}\,.
\end{eqnarray}

The right handed neutrinos may also obtain masses which need not 
have any connection to the inflaton sector. If this be the case then the 
superpotential for the right handed neutrino sector can be written 
as~\cite{Allahverdi,Mazumdar}
\begin{equation}
\label{superpot1}
W=\frac{1}{2}g {\bf X}{\bf N}{\bf N}+h{\bf N}{\bf H}_u {\bf L}+
\frac{1}{2} M_N {\bf N}{\bf N}\,,
\end{equation}
where $g,~h$ are the Yukawas, and $M_{N}$ is the right handed neutrino 
mass term, which breaks the lepton number. We work on a diagonal basis 
for the right handed neutrinos and we assume that the texture is such 
that the lightest right handed neutrino mass is larger than the reheat 
temperature. We call this type of model as type $II$, because the 
masses of the right handed neutrinos are completely independent of 
the inflaton sector.

In the above superpotential note that we have an explicit coupling
between ${\bf X}$ and ${\bf N}$. This coupling is not absolutely
necessary, but $X$ being a SM gauge singlet can couple to the right
handed neutrino sector. The above superpotential has an advantage that
the inflaton can decay via the right handed neutrino sector (off-shell
or on-shell) to the Higgs and the lepton doublet. Therefore reheating
the Universe with the SM degrees of freedom or more precisely twice
the SM relativistic degrees of freedom. Reheating naturally provides
the way out of equilibrium condition.

%%%%%%%%%%%%%%%%%%%%%%%%%%%%%%%%%%%%%%%%%%%%%%%%%%%%%%%%%%%%%%%%%%%

\section{Density perturbations from sneutrino}

In fact one can also imagine that the Yukawa coupling, $g$, in 
Eq.~(\ref{superpot1}) has a non-renormalizable contribution of 
the form~\cite{Mazumdar}
\begin{equation}
\label{coupling}
g=g_{0} \left(1+\frac{\bf N}{M_{p}}+...\right)\,.
\end{equation}
Such a coupling can be easily accommodated at the level of superpotential.

If the sneutrino field is light enough compared to the Hubble expansion
during inflation, e.g. the lightest of the sneutrino field, then the
perturbations generated in the sneutrino field can seed perturbations in
the inflaton sector. The sneutrino fluctuations are isocurvature in nature,
which are converted into the adiabatic fluctuations at the time of reheating. 

Note that in this case the inflaton coupling to the matter field, $g$, is
fluctuating. The reheat temperature, which is given by the inflaton 
coupling to the right handed (s)neutrino field, 
$T_{rh}\sim g\sqrt{m_{X} M_{p}}$, also fluctuates. Since the energy 
density stored in the relativistic species is $\rho_{r}\propto T_{rh}^4$, 
therefore during inflaton dominated oscillations the ratio of energy 
densities at two different times is given by
\begin{equation}
\label{coup1}
\frac{\rho_{2}}{\rho_{1}}=\left(\frac{g_{1}}{g_2}\right)^{4/3}\,.
\end{equation}
The factor $4/3$ appears due to red-shift of the scale factor,
which is during inflaton oscillations following $a(t)\propto t^{2/3}$
\cite{Dvali,Enqvist,Postma}. This gives rise to the fluctuations in 
the energy density which is finally imprinted upon CMB. The fluctuation 
in the energy density of the relativistic species is given 
by \cite{Dvali,Enqvist,Postma}
\begin{equation}
\label{coup2}
\frac{\delta \rho}{\rho} =-\frac{4}{3}\frac{\delta g}{g}=-\frac{4}{3}
\frac{\delta N}{N}\,.
\end{equation} 
Yet another useful way of imagining the coupling term $g$ in 
Eq.~(\ref{coupling}) as a fluctuating mass term for the 
inflaton $X$. 

%%%%%%%%%%%%%%%%%%%%%%%%%%%%%%%%%%%%%%%%%%%%%%%%%%%%%%%%%
\subsection{Multi-field perturbations}

The perturbations are defined on a finite energy density hypersurface 
foliated in a coordinate system such that the metric perturbation is
$\zeta$, and the metric (for a detailed discussion on cosmological
density perturbations, see~\cite{Brandenberger}) is given by
\begin{equation}
ds^2= a^2(t)\left(1+2\zeta \right)\delta_{ij}dx^{i}dx^{j}\,,
\end{equation}
where $a$ is the scale factor. The time evolution of the curvature 
perturbation, $\zeta$, on scales larger than the size of the horizon 
is given by~\cite{Bardeen,Kodama,Gordon}
\begin{equation}
\label{curv0}
\dot\zeta=-\frac{H}{\rho+P}\delta P_{{\rm nad}}\,,
\end{equation}
where $P_{{\rm nad}}\equiv \delta P-c_{s}^2\delta\rho$ is the non-adiabatic
pressure perturbation. The adiabatic sound speed is 
$c_{s}^2=\dot P/\dot\rho$, where $P$ and $\rho$ are the total pressure
and the energy density. For a single field $\delta P_{{\rm nad}}=0$, 
therefore on large scales the curvature perturbation is pure adiabatic 
in nature with $\zeta={\rm constant}$. This is not true in presence of many
fields, because the relative pressure perturbations between fields can give
non-zero contribution to $\delta P_{\rm nad}$.

The total curvature perturbation Eq.~(\ref{curv0}), for many fields, can 
be written in terms of various components~\cite{Kodama,Gordon}
\begin{equation}
\label{curv1}
\zeta=\sum_{\alpha}\frac{\dot\rho_{\alpha}}{\dot\rho}\zeta_{\alpha}\,.
\end{equation}
Isocurvature or entropy perturbations describe the difference between 
the curvature perturbations \cite{Kodama} 
\begin{equation}
\label{iso}
S_{\alpha \beta} =  3\left(\zeta_{\alpha}-\zeta_{\beta}\right)=-3 H \left( 
\frac{\delta \rho_\alpha}{\dot{\rho}_\alpha}
-\frac{\delta \rho_\beta}{\dot{\rho}_\beta}
\right)\,.
\end{equation}
With the help of Eqs.~(\ref{curv1},\ref{iso}), we obtain a 
useful relationship connecting the curvature and the entropy perturbations
\begin{equation}
\zeta_{\alpha}=\zeta+\frac{1}{3}\sum_{\beta}\frac{\dot\rho_{\beta}}{\dot\rho}
S_{\alpha\beta}\,.
\end{equation}
An important quantity is the gauge invariant comoving curvature perturbations
which is defined as 
\begin{equation}
\label{comcurv}
{\cal R}=H\sum_{\alpha}\left(\frac{\dot\phi_{\alpha}}{\sum_{\beta}\dot
\phi_{\beta}^2}\right)Q_{\alpha}\,,
\end{equation}
where $Q_{\alpha}$ is the Sasaki-Mukhanov variable defined in terms of 
the gauge invariant quantities
\begin{equation}
Q_{\alpha}\equiv \delta\phi_{i}+\frac{\dot\phi_{i}}{H}\psi\,.
\end{equation}
where $\psi$ is related to the curvature perturbations by
\begin{equation}
\psi=-\zeta+H\frac{\delta\rho}{\dot\rho}\,.
\end{equation}
The comoving curvature perturbation is defined in terms of the 
Sasaki-Mukhanov variable~\cite{Gordon}
\begin{equation}
{\cal R}_{\alpha}\equiv \psi+\frac{H}{\dot\phi_{\alpha}}Q_{\alpha}\,.
\end{equation} 
The comoving curvature perturbation is dominated by the field 
with a dominating kinetic term. 

There is another useful gauge invariant combination
\cite{Kodama}
\begin{equation}
\delta_{\alpha\beta}\equiv \left(\frac{\delta\phi_{\alpha}}{\dot\phi_{\alpha}}
-\frac{\delta\phi_{\beta}}{\dot\phi_{\beta}}\right)\,,
\end{equation}
and the isocurvature perturbations can be defined as~\cite{Hwang} 
\begin{equation}
\label{entropy}
S_{\alpha\beta}=a^3 \frac{d}{dt}\left(\frac{\delta_{\alpha\beta}}{a^3}\right)
\,.
\end{equation}
We will use the above results in the coming sections.

Especially for two fields case it is fairly easy to investigate the
adiabatic and the isocurvature fluctuations~\cite{Polarski,Gordon}. 
Our two fields are the inflaton and the sneutrino field. For the 
purpose of illustration, we assume there is a single sneutrino 
component responsible for the fluctuations, though in principle all 
three sneutrino components could feel the fluctuations. However the 
lightest among all will have a greater impact, or one can also assume 
that all three neutrinos are nearly degenerate. In which case a linear 
combination of all the species provide the entire perturbations.

We define the adiabatic component as $\sigma$ and the entropic component 
by $s$, such that,
\begin{eqnarray}
\label{decom1}
\delta\sigma&=&(\cos\theta)\delta X +(\sin\theta)\delta \widetilde N\,,\\
\label{decomp2}
\delta s &=&(\cos\theta)\delta \widetilde N-(\sin\theta)\delta X\,,
\end{eqnarray}
where
\begin{equation}
\cos\theta={\dot X}/{\sqrt{\dot X^2+\dot{\widetilde N}^2}}\,,~~
\sin\theta={\dot{\widetilde N}}/{\sqrt{\dot X^2+\dot{\widetilde N}^2}}\,.
\end{equation}
Therefore 
\begin{eqnarray}
\label{entropic}
\delta s =\frac{\dot X\dot{\widetilde N}}
{\sqrt{\dot X^2+\dot{\widetilde N^2}}}\left(\frac{\delta{\widetilde N}}
{\dot{\widetilde N}}-\frac{\delta X}{\dot X}\right)\, 
=\frac{\dot X\dot{\widetilde N}}{\sqrt{\dot X^2+\dot{\widetilde N^2}}}
\delta_{{\widetilde N}X}\,.
\end{eqnarray} 
The comoving perturbations, ${\cal R}$, can be calculated from 
Eq.~(\ref{decom1}) in a spatially flat gauge where $\psi=0$,
\begin{equation}
\label{main1}
{\cal R}\approx H\left(\frac{\dot X\delta X+\dot{\widetilde N}
\delta{\widetilde N}}
{\dot X^2+\dot{\widetilde N}^2}\right)=H\frac{\delta \sigma}{\dot\sigma}\,.
\end{equation} 
Here we assumed slow roll conditions. The entropy perturbations 
can be calculated by combining Eqs.~(\ref{entropy},\ref{entropic}).
\begin{equation}
\label{main2}
S =-H\left(\frac{\delta{\widetilde N}}{\dot{\widetilde N}}-\frac{\delta X}
{\dot X}\right)=-H\frac{\sqrt{\dot X^2+\dot{\widetilde N^2}}}
{\dot X\dot{\widetilde N}}\delta s\,.
\end{equation} 
For $\dot X\gg \dot{\widetilde N}$, the above expression reduces to
$S\approx H(\delta S/\dot{\widetilde N})$, which has some significance when 
dealing with the isocurvature fluctuations. Also note that when the 
perturbations in the inflaton is assumed to be small, such that
$\delta X\ll \delta{\widetilde N}$, or $\dot X\gg \delta X$, then 
the entire perturbations come from the sneutrino field, e.g. 
$S\approx-H(\delta{\widetilde N}/\dot{\widetilde N})$.
Note that when the perturbations in the inflaton $X$ is neglected
then the entropy perturbation arises from the sneutrino sector which 
is solely responsible for feeding the adiabatic mode. This is a
special case which we will discuss later on.

Due to the random Gaussian vacuum fluctuations of $X,~\widetilde N$, 
the fields acquire a spectrum at the time of horizon crossing 
\begin{equation}
\left.{\cal P}_{\delta X}\right|_{\ast}\approx
\left.{\cal P}_{\delta {\widetilde N}}\right|_{\ast}\approx 
\left(\frac{H_{\ast}}{2\pi}\right)^2\,,
\end{equation} 
with zero cross correlation, e.g. 
${\cal C}_{\delta X,\delta\widetilde N}=0$, which we assume for 
simplicity. In terms of local rotations, $\delta s,~\delta\sigma$, 
the spectrum is also proportional to $(H_{\ast}/2\pi)^2$. Therefore
\begin{equation}
\label{spectrum1}
\left.{\cal P}_{R}\right|_{\ast}\approx \left.\left(\frac{H^2}{2\pi\dot\sigma}
\right)^2\right|_{\ast}\sim \frac{8}{3\epsilon}\frac{V_{\ast}}{M_{p}^4}\,.
\end{equation}
where $\epsilon=\epsilon_{X}+\epsilon_{\widetilde N}$. Similarly for 
the isocurvature fluctuations the spectrum is given by
\begin{equation}
\label{spectrum2}
\left.{\cal P}_{S}\right|_{\ast}\approx \left.\left(\frac{\dot\sigma H^2}
{2\pi\dot X\dot{\widetilde N}}\right)^2\right|_{\ast}\,.
\end{equation}
This expression reduces to a simple form if $\dot X\gg\dot{\widetilde N}$, 
then $\left.{\cal P}_{S}\right|_{\ast}\approx \left.\left({H^2}/{2\pi\dot{\widetilde N}}\right)^2\right|_{\ast}$.

%%%%%%%%%%%%%%%%%%%%%%%%%%%%%%%%%%%%%%%%%%%%%%%%%%%%%%%%%%%%%%

\section{Some Examples}

\subsection{Isocurvature fluctuations in type $I$ scenarios}

During inflation the sneutrino condensate evolves in the non-renormalizable
potential given by Eq.~(\ref{source}). In addition, we assume that 
there is no Hubble-induced mass term for the sneutrino even after inflation, 
so that the field will continue slow-rolling in the non-renormalizable 
potential down to the amplitude ${\widetilde N}_{osc}$, when its 
energy density is determined by
$V({\widetilde N}_{osc})\sim M_{N}^2{\widetilde N}_{osc}^2$. 
In general, the equations of motion for the homogeneous and the 
fluctuation parts are written by
\begin{eqnarray}
\label{eqm00}
    & & \ddot{\widetilde N}+3H\dot{\widetilde N}+V'(\widetilde N) = 0\,, \\
\label{eqm01}
    & & \delta\ddot{\widetilde N}_k+3H\delta\dot{\widetilde N}_k
    + \frac{k^2}{a^2}\delta{\widetilde N}_k+V''(\widetilde N) 
\delta{\widetilde N}_k = 0\,.
\end{eqnarray}
Since we are interested only in the super horizon mode ($k\rightarrow 0$), 
using the slow roll approximations we have
\begin{eqnarray}
    \label{sr-homo}
    & & 3H\dot{{\widetilde N}}+V'({\widetilde N}) = 0\,, \\
    \label{sr-fl}
    & & 3H\delta\dot{{\widetilde N}}+V''({\widetilde N}) 
\delta{\widetilde N} = 0\,.
\end{eqnarray}
Hereafter we omit the subscript $k$, understanding that $\delta{\widetilde N}$
is for the super horizon mode. Then it is easy to obtain the evolution
of the ratio of the fluctuation and the homogeneous mode in a                 
$V_{NR} \propto {\widetilde N}^{6}$ potential. The result is
\cite{Enqvist}
\begin{equation}
    \frac{\delta{\widetilde N}}{{\widetilde N}} \sim 
    \left(\frac{\delta{\widetilde N}}{{\widetilde N}}\right)_i
    \left(\frac{{\widetilde N}}{{\widetilde N}_i}\right)^{4}\,,
\end{equation}
where $i$ denotes the initial values.

During inflation the homogeneous field obeys Eq. (\ref{sr-homo}), 
which can be easily integrated to yield
\begin{equation}
    \frac{{\widetilde N}}{{\widetilde N}_i} \simeq \left( 1 + 
      \frac{1}{15} \ \frac{V''({\widetilde N}_i)}{H^2}{\cal N}_{e}
      \right)^{-\frac{1}{4}}\,.
\end{equation}
Since we are concerned within a slow-roll regime, it is 
reasonable to require $V''({\widetilde N}_i)/H^2 \lesssim 1$. 
Hence we have, ${\widetilde N}/{\widetilde N}_i \approx 0.67$,
for the last $60$ e-folds in this case. This implies that
the amplitude of the fluctuation relative to its homogeneous part
decreases only by a factor $\simeq 0.2$. Hence during this stage 
there is less damping. Notice that slower the condensate field
rolls during the last $60$ e-folds, the less damping there is.

After inflation the sneutrino condensate slow-rolls (albeit marginally), 
i.e., $V''({\widetilde N}) \sim H^2$, and we can still use the 
slow-roll approximations, e.g. Eqs.~(\ref{sr-homo},\ref{sr-fl}). 
This will give the largest estimate on the dilution of the amplitude.
During this stage the field amplitude is given by 
${\widetilde N}\sim (H M_{p}/\lambda)^{1/2}$, while the 
Hubble parameter changes from $H_*$ to $M_{{\widetilde N}}$. 
As a consequence, there is a damping given by
\begin{equation}
    \frac{\left(\frac{\delta{\widetilde N}}{{\widetilde N}}\right)_{osc}}
    {\left(\frac{\delta{\widetilde N}}{{\widetilde N}}\right)_*} \sim 
    \left(\frac{M_{{\widetilde N}}}{H_*}\right)^2\sim 3\times g^2\,.
\end{equation}
The last equality follows from Eq.~(\ref{type1}).
 
At the time when the sneutrino decays, e.g. 
$H\equiv\Gamma_{N}\sim h^2M_{N}/4\pi$, where the largest Yukawa 
coupling is that of the tau doublet, of the order of 
$h\sim {\cal O}(10^{-4})$, which takes place when the net lepton 
number is given by $n_{L}=\rho_{\widetilde N}/M_{\widetilde N}\sim M_{\widetilde N}{\widetilde N}^2$.
Therefore the isocurvature perturbations are given by
\begin{eqnarray}
\label{isoc1}
S&=&\frac{\delta n_{L}}{n_{L}}\frac{\Omega_{B}}{\Omega_{m}}=
\frac{\delta n_{B}}{n_{B}}\frac{\Omega_{B}}{\Omega_{m}}\,\nonumber \\
&\approx &
6g^2\left(\frac{\delta {\widetilde N}}{\widetilde N}\right)_{\ast}
\frac{\Omega_{B}}{\Omega_{m}}\approx 6g^2\frac{H_{\ast}}
{2\pi{\widetilde N}_{\ast}}\frac{\Omega_{B}}{\Omega_{m}}\,.
\end{eqnarray}
The subscript $m$ denotes total matter density.
To compare the two types of fluctuations, it may be useful to consider
the ratio between the adiabatic and the isocurvature 
\begin{eqnarray}
\alpha\sim\left.\frac{{\cal P}_{S}^{1/2}}{{\cal P}_{R}^{1/2}}\right|_{\ast}
&\approx&\frac{3g^2\lambda^2}{32\sqrt{2}\pi^3}\frac{M_{p}^2}{X_{\ast}
{\widetilde N}_{\ast}}\left(\frac{\Omega_{B}}{\Omega_{m}}\right)\,,
\nonumber \\
&\approx&\frac{3g^2\lambda}{16\sqrt{2}\pi^2\sqrt{{\cal N}_{e}}}
\frac{M_{p}}{{\widetilde N}_{\ast}}
\left(\frac{\Omega_{B}}{\Omega_{m}}\right)\,.
\end{eqnarray}
The last equality comes due to Eq.~(\ref{lim}). There are 
couple of points to be mentioned. First of all in type I model, 
we assumed that the major adiabatic fluctuations arose from the 
inflaton sector. The total power spectrum is given by 
${\cal P}={\cal P}_{R}+{\cal P}_{S}$. We also assumed that there
is no correlation. Recent observations from WMAP data constraints this 
ratio~\cite{WMAP}. The uncorrelated isocurvature fluctuations in the CDM 
has been presented in a recent analysis, which suggests the ratio 
$\alpha<0.31$ at $95\%$ c.l. If we take 
${\cal N}_{e}=50$, $\Omega_{B}h^2=0.023$, $\Omega_{m}h^2=0.133$, we find 
\begin{equation}
\label{boundsuno}
{\widetilde N}_{\ast}> 10^{-3}g^2\lambda M_{p}\,.
\end{equation}
On the other hand, in order to have inflation and the spectral index 
governed solely by the inflaton field, ${\widetilde N}<{\widetilde N}_{c}$, 
from Eq.~(\ref{critical}), which puts additional constraints on model
parameters such as $g,~\lambda,~\lambda_{1},~v_{\phi}$, or in terms of the
right handed neutrino mass scale $M_{N}$, the constraints translate to
\begin{equation}
\label{boundsdue}
M_{N}\geq 5\times10^{-10}g^7\lambda^2\lambda_{1}M_{p}\,.
\end{equation}
Note that the scale of the right handed neutrino is quite sensitive 
to the Yukawa coupling $g$. For the couplings order one, we obtain a 
reasonable bound $M_{N}\geq 5\times 10^{8}$~GeV. Also note that if 
we assume that the right handed neutrino sector is embedded in a 
gauge sector such as GUT, the ${\widetilde N}_{\ast}\leq 10^{16.5}$~GeV,
assuming that the GUT scale is at $10^{16.5}$~GeV, suggesting that the 
Yukawas, $g^2\lambda \ll {\cal O}(1)$. Additional constraints will resurface 
based on thermal history of the Universe during inflaton decay, 
e.g. sneutrino interacting in a finite temperature thermal bath.
Some of these issues are model dependent, such as how 
inflaton is reheating the Universe, whether the inflaton is decaying 
predominantly into the MSSM degrees of freedom or some other, etc. 
Hopefully we will address them in a future publication.
Strictly speaking the bounds Eqs.~(\ref{boundsuno},\ref{boundsdue})
hold true only if the finite temperature effects are negligible.

%%%%%%%%%%%%%%%%%%%%%%%%%%%%%%%%%%%%%%%%%%%%%%%%%%%%%%%%%%%%%%%%%%%%%
\subsection{Isocurvature fluctuations in type $II$ scenarios}

In type $II$ scenario there is no need for the explicit inflaton coupling 
to the right handed neutrino sector. In fact if $g=0$ in 
Eq.~(\ref{superpot1}), then the sneutrino evolution will be similar to
what we already discussed in our previous scenario. The 
fluctuations in the inflaton and the sneutrino sector could be 
treated separately on scales larger than the size of the horizon. 
Even in this case one can expect non-renormalizable superpotential 
contribution to the sneutrino. Nevertheless if the initial amplitude 
for the sneutrino, e.g. 
${\widetilde N}_{i}\leq (M_{N}M_{p}/2\lambda_1)^{1/2}$, then the 
non-renormalizable contribution will not play any significant role,
and therefore there will be no damping in the amplitude 
of the fluctuations in the sneutrino sector after the end of 
inflation. The above estimation for the baryon isocurvature 
fluctuations in Eq.~(\ref{isoc1}) holds true without 
the damping factor $6g^2$. As a result the sneutrino vev must have 
\begin{equation}
{\widetilde N}_{\ast}>10^{-4}\lambda M_{p}\,,
\end{equation}
assuming that the inflaton sector is given by $V(X)$, see  
Eq.~(\ref{source}). The bound on the neutrino mass scale  
arises from the dominance of the inflaton energy density,
\begin{equation}
M_{N}< 10^{4}\frac{v_{\phi}^2}{M_{p}}\,,
\end{equation}
over the sneutrino condensate.

The non vanishing inflaton right handed neutrino coupling gives rise to 
a completely new feature. The isocurvature fluctuations in this case 
will be certainly correlated, and there is a new possibility that the 
sneutrino induced isocurvature fluctuations get converted into the 
adiabatic fluctuations. We will discuss this issue later on in a separate 
subsection. 

%%%%%%%%%%%%%%%%%%%%%%%%%%%%%%%%%%%%%%%%%%%%%%%%%%%%%%%%%%%%%%%%%%%%

\subsection{What if $M_{N}\geq H_{inf}$~?}

It is quite possible that the right handed neutrino mass scale is greater 
than or equal to the Hubble expansion during inflation. In either case the
sneutrino will roll down to its minimum from its initial vev in less than 
one Hubble time and starts oscillating before settling down with a vanishing
kinetic term. If the Yukawa coupling $h$ is sufficiently large, 
$\Gamma_{N}\gg H_{inf}\gg \Gamma_{X}$, then the sneutrino might even decay
during inflation. In an opposite limit $\Gamma_{N}\leq H_{inf}$, the 
sneutrino survives inflation, and feels the quantum fluctuations.

The solutions for Eq.~(\ref{eqm01}) with 
$V({\widetilde N})=\alpha H_{inf}^2{\widetilde N}^2$ is well known
\cite{Liddle} (from here onwards we drop the subscript $inf$)
\begin{equation}
\delta{\widetilde N}_{k}\approx H(Ha)^{-3/2}\left(\frac{k}{aH}\right)^
{-\sqrt{9/4 -\alpha}}\,,
\end{equation}
and the power spectrum follows
\begin{equation}
{\cal P}_{\delta{\widetilde N}}\propto H^{3/2}\left(\frac{k}{aH}\right)^
{3/2-{\rm Re}(\sqrt{9/4-\alpha}~)}\,.
\end{equation}
For $\alpha\gg 9/4$ the real part of the exponent vanishes leaving a 
very steep spectrum for the isocurvature perturbations, e.g. 
${\cal P}_{S}\propto k^{3}$, which is exponentially suppressed by 
the end of inflation. This regime certainly has not much of
interest. However if $\alpha\leq 9/4$ then 
${\cal P}_{S}\propto k^{2\alpha/3}$~\cite{Liddle}. Though in this 
case the tilt in the isocurvature power spectrum plays an important
role in constraining
\begin{equation}
n_{iso}=\frac{4\alpha}{3}\,.
\end{equation}
For the uncorrelated isocurvature fluctuations, the constraint on the
isocurvature spectral index is $n_{iso}=1.02$ at $95\%$ c.l.~\cite{Bellido}. 
This gives a limit on the right handed neutrino mass scale, $M_{N}\sim 0.7 H$.

%%%%%%%%%%%%%%%%%%%%%%%%%%%%%%%%%%%%%%%%%%%%%%%%%%%%%%%%%%%%%%%%%%%
\subsection{Correlated baryon-Isocurvature fluctuations}

Most of the examples belong to the category where the isocurvature and
the adiabatic fluctuations are uncorrelated, because the two fields
had fluctuations independent of each other 
$\langle R_{\ast}S_{\ast}\rangle =0$. Here we consider a simple example
$\langle R_{\ast}S_{\ast}\rangle\neq 0$. We already set up a
superpotential term Eq.~(\ref{superpot1}) with a coupling Eq.~(\ref{coupling}).

Following Eqs.~(\ref{coup1},~\ref{coup2}), we notice that the fluctuation
in the reheat temperature is given by
\begin{equation}
\label{rel}
\frac{\delta T_{rh}}{T_{rh}}=-\frac{1}{3}\frac{\delta g}{g}\sim 
-\frac{\delta \widetilde N_{\ast}}{3M_{p}}\sim -\frac{H_{\ast}}{6\pi M_{p}}\,.
\end{equation}
An important point to note is that the baryon asymmetry is also 
proportional to $g$, see Ref.~\cite{Mazumdar}. Therefore baryons 
also feel the spatial fluctuations.
\begin{equation}
\label{excite}
\frac{\delta \eta_{B}}{\eta_{B}}\sim -\frac{1}{3}\frac{\delta g}{g}\sim 
-\frac{\delta\widetilde N}{3M_{\rm p}}\sim 
\frac{\delta T_{\rm R}}{T_{\rm R}}\neq 0\,.
\end{equation}
The origin of $-1/3$ factor has a similar origin as Eq.~(\ref{rel}). 
Note that the fluctuations in the baryon asymmetry is proportional 
to the fluctuations in the inflaton coupling, and therefore fluctuations 
in the reheat temperature. This shows that the baryonic asymmetry 
does not follow the adiabatic density perturbations, instead the
perturbation in the baryons is correlated baryon-isocurvature in nature.  
The two fluctuations; isocurvature and adiabatic perturbations are not 
independent of each other. Rather the former feeds the latter ones.

The baryon-isocurvature fluctuations leaves its imprint 
upon the cosmic micro wave background radiation.  
\begin{eqnarray}
S_{B}&=&\frac{\delta\eta_{B}}{\eta_{B}}=\frac{\delta T_{rh}}{T_{rh}}\,,\\
\zeta&=&-H\frac{\delta\rho_{\gamma}}{\dot\rho_{\gamma}}=\frac{1}{4}
\frac{\delta\rho_{\gamma}}{\rho_{\gamma}}=\frac{\delta T_{rh}}{T_{rh}}\,,
\end{eqnarray}
where the subscript $\gamma$ denotes MS(SM) radiation. Therefore 
we find $|S_{B}/\zeta|=1$. This toy model has a unique prediction which can
be ruled out easily from the future cosmic microwave background experiments.

%%%%%%%%%%%%%%%%%%%%%%%%%%%%%%%%%%%%%%%%%%%%%%%%%%%%%%%%%%%%%%%%%%%
\section{Conclusion}  

We argued that the non-thermal leptogenesis is potentially testable from 
its contribution to the baryon isocurvature fluctuations. There could be 
many other sources for the isocurvature perturbations during inflation 
including the most competitive candidate ``cold dark matter''. However 
within SUSY excellent conditions arise naturally for a thermal production
of the CDM. We also note that the thermal leptogenesis is a viable 
scheme, nevertheless, it is quite natural that during inflation the sneutrino 
condensate can be created. If the condensate survives inflation and the 
thermal bath created by the inflaton decay products, then the sneutrino 
decay can generate the lepton asymmetry. The asymmetry created in the lepton 
sector inherits the spatial fluctuations from the sneutrinos during 
inflation, which are isocurvature in nature. These isocurvature fluctuations 
could be uncorrelated and/or correlated in nature. In this paper we 
have provided examples which are tied up with the inflation sector, 
and we have also given examples of correlated and uncorrelated 
isocurvature fluctuations. We estimated the ratio $S/R$, which constrains 
various model parameters and also the mass scale of the right handed 
neutrinos. In the simplest realization of type $I$ leptogenesis, 
we found the lightest sneutrino vev to be 
${\widetilde N}>10^{-3}g^2\lambda M_{p}$, and 
$M_{N}\geq5\times10^{-10}g^7\lambda^2\lambda_{1}M_{p}$. In type $II$ case,
${\widetilde N}_{\ast}>10^{-4}\lambda M_{p}$, and 
$M_{N}<10^{4}v_{\phi}^2/M_{p}$. We also noticed that when the sneutrino 
mass is heavier than the Hubble expansion during inflation, then the 
isocurvature perturbations die away. However if their mass ranges 
are close to the Hubble expansion during inflation,
then the important constraint arises from the spectral tilt 
$n_{iso}=(4M^2_{N}/3H_{\ast}^2)$. We also gave an example of a toy model 
where the perturbations are correlated and there is a unique prediction  
for $S/R=1$.

Though, for simplicity we restricted our perturbation analysis to 
the two fields, in reality an elaborate treatment of the perturbations 
of all three generations of the sneutrinos is necessary, nevertheless, 
we catch an interesting glimpse of the problem. In principle the 
formalism developed here can be carried on to incorporating all three
generations, which we leave for future investigation. 
                            
%%%%%%%%%%%%%%%%%%%%%%%%%%%%%%%%%%%%%%%%%%%%%%%%%%%%%%%%%%%%%%%%%%%%%%%% 
\vskip5pt

%\section*{Acknowledgments}
A. M. is a CITA National fellow. He is thankful to Rouzbeh
Allahverdi, Katlai Balaji, Robert Brandenberger, Zurab Berezhiani, 
Guy Moore and Marieke Postma for valuable discussion.

%%%%%%%%%%%%%%%%%%%%%%%%%%%%%%%%%%%%%%%%%%%%%%%%%%%%%%%%%%%%

%\vskip-22pt

\end{document}